\documentclass[12pt]{article}
\usepackage{amsmath,amssymb}

\usepackage{graphicx}
\usepackage{wrapfig}

\usepackage{hyperref}
\usepackage{cite}

\def\beq{\begin{eqnarray}}
\def\eeq{\end{eqnarray}}

\def\la{\langle }
\def\ra{\rangle }

\newcommand{\be}{\begin{equation}}
\newcommand{\ee}{\end{equation}}
\newcommand{\bea}{\begin{eqnarray}}
\newcommand{\eea}{\end{eqnarray}}
\newcommand{\bg}{\begin{gather}}

\newcommand{\bseq}{\begin{subequations}}
\newcommand{\eseq}{\end{subequations}}

\def\be{\begin{eqnarray}}
\def\ee{\end{eqnarray}}
\def\lb{\label}

\begin{document}

\title{\textbf{Conformal a-charge, correlation functions  \\
and conical defects}}

\vspace{2cm}
\author{ \textbf{ Sergey N. Solodukhin }} 

\date{}
\maketitle
\begin{center}
  \hspace{-0mm}
  \emph{  Laboratoire de Math\'ematiques et Physique Th\'eorique  CNRS-UMR
7350 }\\
  \emph{F\'ed\'eration Denis Poisson, Universit\'e Fran\c cois-Rabelais Tours,  }\\
  \emph{Parc de Grandmont, 37200 Tours, France}
\end{center}

{\vspace{-11cm}
\begin{flushright}
\end{flushright}
\vspace{11cm}
}



\begin{abstract}
\noindent { In this note we demonstrate that, as we conjectured earlier in \cite{SS}, the  a-charge in the conformal anomaly in dimension $d=2n$ manifests 
in a $n$-point correlation function of energy momentum tensor  of a CFT considered in flat spacetime with a conical defect.
We consider in  detail  dimensions $d=2,\, 4,\, 6$ and give a general formula for arbitrary $n$.

}
\end{abstract}

\vskip 2 cm
\noindent
\rule{7.7 cm}{.5 pt}\\
\noindent 
\noindent
\noindent ~~~ {\footnotesize e-mail:  Sergey.Solodukhin@lmpt.univ-tours.fr}

\newpage
    \tableofcontents
\pagebreak

\newpage

\section{ Introduction}
\setcounter{equation}0
Conformal symmetry plays an increasingly important role in the contemporary  theoretical models.
Although always explicitly broken in Minkowski spacetime by presence of massive particles and dimensionful
couplings this symmetry restores at some critical points of the theory. These points are of special attention 
since it is believed that the RG evolution of the theory between the critical points is irreversible.
The complete description of this irreversibility is  one of the actively discussed important problems.  

The other way to break the conformal symmetry is to place the theory on a curved background.
Then, the otherwise traceless quantum stress energy tensor acquires a non-trivial trace which, in general even dimension $d=2n$,
can be represented in terms of certain curvature invariants. One of these invariants is the Euler density,
the quantity which being integrated over the whole manifold produces a topological invariant, the Euler number.
The other quantities are invariants under conformal  transformations. They are certain polynomials of the Weyl tensor and its derivatives.
All these terms represent the conformal anomaly first discovered by M. Duff and  D. Capper  in 1974 \cite{Capper:1974ic}.
The contribution due to topological Euler density has been later called the anomaly of type A (or the a-charge) while the terms constructed by means of the Weyl tensor
represent anomaly of type B in the terminology of \cite{Deser:1993yx}.  
The conformal anomalies appear in any even dimension $d=2n$.

Much attention in the recent studies has been payed to the anomaly of type A. This is mostly due to its topological nature and
the conjectured monotonic behavior of the a-charge during the RG flow   \cite{Cardy:1988cwa}. This conjecture has been advanced  to the level of a theorem
by recent works \cite{Komargodski:2011vj}.

In flat spacetime  the conformal anomaly of type A manifests itself in higher point correlation functions of energy momentum tensor. Thus, in dimension $d=2$
it shows up in 2-point correlation function $\la TT\ra$  \cite{Deser:1993yx}, in dimension $d=4$ in 3-point function $\la TTT\ra$ \cite{Osborn:1993cr} and so on.

On the other hand, in Minkowski spacetime, the conformal a-anomaly manifests   in the logarithmic term in  entanglement entropy calculated 
if the entangling surface is round sphere \cite{Solodukhin:2008dh}.
 This suggests that the monotonic nature of RG flow could be analyzed
entirely in terms of entanglement entropy \cite{SS} (earlier work on relation of the a-theorem and entanglement entropy includes \cite{EC}).  A useful technical tool to study entanglement entropy is to introduce  a small angle deficit
at the entangling surface $\Sigma$ (see \cite{Callan:1994py} and for a review see \cite{EE}). The entropy then can be calculated as a response of the quantum field in question to this conical singularity.

Effectively, a conical defect in otherwise flat spacetime reduces the dimensionality of the problem. 
In particular, it was conjectured in \cite{SS} that in $d=4$ the conformal a-anomaly should be visible already in 2-point function
$\la TT\ra$ considered in flat spacetime with a conical defect.  This conjecture  caused a certain disbelief in the CFT community.
In this note we prove it using, in particular,  the recently proposed  \cite{Smolkin:2014hba}  correspondence between the $N$-point correlation functions
in spacetime with a conical defect and certain $(N+1)$-correlation functions in Minkowski spacetime without defects.
In fact, we can now prove a  more general statement that the  a-charge in the conformal anomaly in dimension $d=2n$ manifests
in a $n$-point correlation function of the  energy momentum tensor considered on spacetime with a conical defect. More specifically, we show this for dimensions
$d=2,\, 4, \, 6$ and give a general formula for arbitrary $n$.

\section{ The tools}
\setcounter{equation}0
Before proceeding with our analysis we pause here to explain the technical tools to be used.

\subsection{Curvature invariants of conical space}
We shall use the distributional nature of the conical singularity.
Due to this nature in the presence of a conical singularity the curvature 
has a delta-like contribution at the singular surface \cite{Fursaev:1995ef}
\begin{eqnarray}
&&R^{\mu\nu}_{\ \ \alpha\beta} = \bar{R}^{\mu\nu}_{\ \
\alpha\beta}+
2\pi (1-\alpha) \left( (n^\mu n_\alpha)(n^\nu n_\beta)- (n^\mu
n_\beta)
(n^\nu n_\alpha) \right) \delta_\Sigma \, ,\nonumber \\
&&R^{\mu}_{ \ \nu} = \bar{R}^{\mu}_{ \ \nu}+2\pi(1-\alpha)(n^\mu
n_\nu)
\delta_\Sigma \, , \nonumber       \\
&&R = \bar{R}+4\pi(1-\alpha) \delta_\Sigma\, ,
\label{singular-curvature}
\end{eqnarray}
where
$n^\mu_k \, , \, k=1,\, 2$ are two orthonormal
vectors
orthogonal to the surface $\Sigma$, $(n_\mu n_\nu)=\sum_{k=1}^{2}n^k_\mu
n^k_\nu$,
and the quantity $\bar{\cal R}$ is the regular part of the curvature.

\subsection{Minkowski/conical defect duality}
The second important ingredient in our work is to use the recently proposed in \cite{Smolkin:2014hba} correspondence between Minkowski spacetime and spacetime with
a conical defect. According to this correspondence we have a relation
\be
{\cal P}\la {\cal O}_1(x_1)...{\cal O}_N(x_N)\ra_{{\cal C}_\alpha}=\la {\cal O}_1(x_1)...{\cal O}_N(x_N)K_0\ra
\, ,
\lb{2}
\ee
where operator ${\cal P}=-\lim_{\alpha\rightarrow 1}\, \frac{\partial}{\partial\alpha}$, the correlation function in the left hand side is calculated on spacetime with a defect with angle deficit $\delta=2\pi (1-\alpha)$
and in the right hand side the correlation function is computed in Minkowski spacetime without any defects. 
The operator $K_0$ is the modular Hamiltonian. For a planar surface it takes the form
\be
K_0=-2\pi\int d^{d-2}y \int_0^\infty dx_1 x_1 T_{22}(x_1,x_2=0,y)\, ,
\lb{1}
\ee
where $(x_1,x_2)$ are Cartesian coordinates in the transverse space, $\Sigma$ is located at the origin
$x_1=x_2=0$ and $y^i$ with $i=3,..,d$ are Cartesian coordinates on $\Sigma$. In this notation $x_2$ plays the role of Eulcidean time and $T_{22}$ is the respective component of the energy-momentum tensor.  It is useful to note that the modular Hamiltonian  generates angular evolution  in plane $(x_1,x_2)$.
Relation (\ref{2}) associates the leading in $(1-\alpha)$ term in the correlation function $\la ..\ra_{{\cal C}_\alpha}$ with a higher-point function in Minkowski spacetime.

\section{ Dimension $d=2$}
\setcounter{equation}0

In two dimensions the conformal anomaly is entirely of the type A,
\be
\la T(x_1,x_2)\ra=\frac{c}{24\pi} R\, ,
\lb{1.1}
\ee
where $T=T^\mu_{\ \mu}$ is trace of vacuum expectation value of energy momentum tensor, $R$ is the Ricci scalar and $c$ is two-dimensional analog of $a$-charge. 
Being considered in flat spacetime this expression vanishes. Therefore, one has to look at a 2-point function \cite{Cappelli:1990yc}, \cite{Deser:1993yx}
\be
\la T(x)T_{\mu\nu}(x')\ra=\frac{c}{12\pi}(\partial_\mu\partial_\nu-\delta_{\mu\nu}\partial^2)\delta^{(2)}(x-x')\, 
\lb{1.2}
\ee
in order to detect the $c$-charge,

According to our proposal in the presence of a conical defect the situation is different and we can see the $c$-charge already in $1$-point function.
In order to prove this we present here two independent derivations
of 1-point correlation function of the CFT energy momentum tensor in spacetime ${\cal C}_\alpha$ with a conical defect. 
This spacetime is everywhere flat except for the conical defect where curvature has a delta-function behavior. 
In the first calculation, we simply apply formula (\ref{singular-curvature}) to the right hand side of (\ref{1.1})
and obtain for the correlation function in space ${\cal C}_\alpha$
\be
\la T^\mu_{\, \mu}(x_1,x_2)\ra_{{\cal C}_\alpha}=(1-\alpha)\frac{c}{6}\, \delta_\Sigma\, ,
\lb{1.3}
\ee
where $\delta_\Sigma=\delta(x_1)\delta(x_2)$.

In the second derivation we use the correspondence (\ref{2}) 
\be
&&\la T(x_1,x_2)K_0\ra=-2\pi\int_0^\infty dx'_1\, x_1'\, \la T(x_1,x_2)T_{22}(x_1',x_2'=0)\ra\nonumber \\
&&=\frac{c}{6}\int_0^\infty dx'_1x_1' \frac{\partial^2}{\partial{{x'_1}^2}}\delta(x_1'-x_1)\delta(x_2)
=\frac{c}{6}\delta(x_1)\delta(x_2)\, ,
\lb{1.4}
\ee
where the 2-point function (\ref{1.2}) has been used, and again reproduce (\ref{1.3}).  We used the relation
\be
\int_0^\infty dx'_1\, x_1' \, \frac{\partial^2}{\partial{{x'_1}^2}}\delta(x_1'-x_1)=\delta(x_1)
\lb{1.5}
\ee
when derived (\ref{1.4}).

\section{ Dimension $d=4$}
\setcounter{equation}0
In four dimensions one has for the trace anomaly
\be
&&\la T\ra=-\frac{a}{64}E_4+\frac{b}{64} W^2\, ,\nonumber \\
&&E_4=R_{\alpha\beta\mu\nu}R^{\alpha\beta\mu\nu}-4R_{\mu\nu}R^{\mu\nu}+R^2\, ,\nonumber \\
&&W^2=R_{\alpha\beta\mu\nu}R^{\alpha\beta\mu\nu}-2R_{\mu\nu}R^{\mu\nu}+\frac{1}{3}R^2\, ,
\lb{2.1}
\ee
where  the second term is  the anomaly of type  $B$, $W$ is the Weyl tensor  and $E_4$ is the Euler density in four dimensions. In this normalization a conformal scalar field
has $a$-charge equal to $1/90\pi^2$. 

Below will shall first focus on the $A$-anomaly and then comment on the irrelevance of the anomaly of type $B$ for the correlation functions we
consider.

As in the two-dimensional case, we present two derivations.

\subsection{Variation of 1-point function}

The first derivation uses the property 
\be
\la T_{\mu\nu}(x)T(x')\ra=\frac{2}{\sqrt{g}} \frac{\delta}{\delta g^{\mu\nu}(x)}\la T(x')\ra\, ,
\lb{2.2}
\ee
where for $1$-point function we take (\ref{2.1}) and consider it on a conical defect so that relations
(\ref{singular-curvature}) should be used. 

Thus, we compute the variation of $E_4$ and then consider this variation on a background of flat space with a conical defect
(the regular part of the curvature in (\ref{singular-curvature}) then  vanishes).
For  the variation under $g^{\mu\nu}\rightarrow g^{\mu\nu}+\delta g^{\mu\nu}$ of the Euler density (\ref{2.1}) in this procedure we find
\be
\delta E_4|_{C_\alpha}=8\pi(1-\alpha)\gamma^{\mu\alpha}\gamma^{\nu\beta}\delta R_{\mu\nu\alpha\beta}\, \delta_\Sigma\, ,
\lb{2.3}
\ee
where $\gamma^{\mu\nu}=g^{\mu\nu}-(n^\mu n^\nu)$ is the induced metric and 
\be
\delta R_{\mu\nu\alpha\beta}=-\frac{1}{2}\left(\partial_\alpha\partial_\nu \delta g_{\mu\beta}+\partial_\beta\partial_\mu \delta g_{\nu\alpha}-\alpha\leftrightarrow \beta\right)
\lb{2.4}
\ee
is variation of the Riemann tensor over flat metric, $\delta g_{\mu\nu}=g_{\mu\alpha}g_{\nu\alpha}\delta g^{\alpha\beta}$.
We notice that this variation of the Euler density is purely intrinsic. Indeed, all derivatives and components of $\delta g^{\mu\nu}$ present in (\ref{2.4}) are along the
surface $\Sigma$. 

Now, using (\ref{2.2}) we arrive at $2$-point correlation function
\be
\la T_{\mu\nu}(x)T(x')\ra_{{\cal C}_\alpha}=\frac{a\pi}{4}(1-\alpha)(\gamma_\mu^{\ \alpha}\gamma_\nu^{\ \beta}-\gamma_{\mu \nu}\gamma^{\alpha\beta})\partial_\alpha\partial_\beta\delta(x-x')\, \delta_\Sigma\, .
\lb{2.5}
\ee
Taking the trace we find
\be
\la T(x)T(x')\ra_{{\cal C}_\alpha}=-\frac{a\pi }{4}(1-\alpha)\partial^2_\Sigma \delta(x-x')\, \delta_\Sigma\, .
\lb{2.6}
\ee
Formulas (\ref{2.5}) and (\ref{2.6}) have appeared earlier in \cite{SS}. 

We notice that these correlation functions are purely intrinsic.
Let us consider the Cartesian coordinates $x=(x_1,x_2,y_i)$, where $y_i$, $i=3,4$ are the coordinates on the surface. Then the induced metric $\gamma_{\mu\nu}$ has the only non-vanishing components $\gamma_{ij}$, $i,j=3,4$. 
Therefore,  (\ref{2.5}) vanishes if at least one of the indexes $(\mu\, \nu)$ takes value $(1,2)$  in the sub-space orthogonal to the surface $\Sigma$.
On the other hand, the derivatives in (\ref{2.5}) and (\ref{2.6}) are acting along the surface $\Sigma$ and the whole correlation function is
supported entirely on the singular surface. 
Taking these comments we can rewrite (\ref{2.5}) in the following form
\be
\la T_{\mu\nu}(x)T(x')\ra_{{\cal C}_\alpha}=\frac{a\pi}{4}(1-\alpha)(\gamma_\mu^{\ i}\gamma_\nu^{\ j}-\gamma_{\mu \nu}\gamma^{ij})\partial_j\partial_j\delta(y-y')\delta(x_1-x_1')\delta (x_2-x_2')\delta_\Sigma\, ,
\lb{2.7}
\ee
where $\delta_\Sigma=\delta(x_1)\delta(x_2).$

These formulas should be compared to those obtained in the case of two-dimensional CFT, see eq.(\ref{1.2}).
They are identical up to the factor $(1-\alpha)$ and delta-functions in orthogonal subspace $(x_1,x_2)$. This observation  gives yet another support to the possible identification, as proposed in \cite{SS}, of a four-dimensional $a$-charge with  $c$-charge
of a two-dimensional CFT defined on a singular surface.

Let us now comment on possible contribution of the anomaly of type $B$, the second term in (\ref{2.1}) proportional to the square of Weyl tensor.
Considering this term on a conical defect we obtain a contribution
\be
W^2|_{{\cal C}_\alpha}=8\pi(1-\alpha)W_{abab}\delta_\Sigma\, ,
\lb{W}
\ee
where $W_{abab}$ is the projection of Weyl tensor on subspace transverse to the surface $\Sigma$. A variation  
$g^{\mu\nu}\frac{\delta}{\delta g^{\mu\nu}}$  of this expression (considered in flat background)  vanishes since Weyl tensor is conformal invariant.
Therefore, the $B$-anomaly does not make any contribution to 2-point function (\ref{2.6}). The latter thus is solely produced by the $a$-charge.

\subsection{Derivation using 3-point function in Minkowski spacetime}
In flat spacetime the $a$-charge of a CFT manifests  in correlation functions of energy momentum tensor starting with $3$-point function.
The exact form of the corresponding contribution in 3-point function has been found by Osborn and Petkou \cite{Osborn:1993cr} (see eq.(8.26) of their paper)\footnote{Notice that energy momentum tensor in 
\cite{Osborn:1993cr} is defined with a minus sign, $T_{\mu\nu}=-\frac{2}{\sqrt{g}}\frac{\delta W}{\delta g^{\mu\nu}}$. This explains the different sign in (\ref{2.8}) relative to (8.26) in \cite{Osborn:1993cr}.} 
\be
&&\la T(x) T_{\sigma\rho}(y) T_{\alpha\beta}(z)\ra=-4\beta_b\, {\cal A}^G_{\sigma\rho,\alpha\beta}(x-y,x-z)+..\, ,\nonumber \\
&&{\cal A}^G_{\sigma\rho,\alpha\beta}(x-y,x-z)=\epsilon_{\sigma\alpha\gamma \kappa}\, \epsilon_{\rho\beta\delta\lambda}\, \partial_\kappa\partial_\lambda(\partial_\gamma \delta(x-y)\partial_\delta \delta(x-z))+\sigma\leftrightarrow\rho \, ,
\lb{2.8}
\ee
where we keep only those terms which are proportional to the $a$-charge and skip all other terms. The exact relation between our $a$ which appears in (\ref{2.1}) and $\beta_b$ is
\be
\beta_b=\frac{a}{64}\, .
\lb{2.9}
\ee

Now we use the correspondence (\ref{2}) with $K_0$ taking the form (\ref{1}) and compute the 2-point function of energy momentum tensor
on a conical defect using (\ref{2.8}). As we from (\ref{1}) the modular Hamiltonian is defined by certain integral of $(22)$ component of energy momentum
tensor. Therefore, we have to first calculate the 3-point function (\ref{2.8}) when $\alpha=\beta=2$.
Then, we have to perform the two integrations contained in definition of $K_0$. One integration is over coordinates $z_3,z_4$ in the orthogonal sub-space
and then the integration over $z_1$.  It is useful to note that in the first integration we get zero for any terms which contain derivatives with respect to variables
$z^i$, $i=3,4$,
\be
\int dz_3dz_4\, \partial_i\partial_{\mu_1}..\partial_{\mu_k}\delta(x-z)=0\, ,\, \, i=3,4
\lb{2.10}
\ee
and $\mu_1$..$\mu_k$ are any indexes from $1$ to $4$.
Therefore, analyzing the 3-point function in question and having in mind that this to be further integrated to get a correlation function with $K_0$ 
one can  neglect all such terms. We shall thus use notation $:=:$ for any equality modulo terms
of the type $\partial_i\partial_{\mu_1}..\partial_{\mu_k}\delta(x-z)$.
We then find that
\be
&&\la T(x) T_{ab}(y) T_{22}(z)\ra :=:\, 0\, , \,\, a,b=1,2\nonumber \\
&&\la T(x) T_{ai}(y) T_{22}(z)\ra :=:\, 0\, ,\,\, a=1,2\, , \,\, i=3,4\nonumber \\
&&\la T(x) T_{ij}(y) T_{22}(z)\ra :=: \, -8\beta_b(\delta_{il}\delta_{kj}-\delta_{ij}\delta_{kl})\partial_l\partial_k\delta(x-y)\partial_1^2\delta(x-z)\, ,
\lb{2.11}
\ee
where in the last line all indexes $i,j,k,l$ take values $3,4$. The two integrations can now be easily performed and we find for the non-vanishing components of the correlation function
\be
\la T(x)T_{ij}(y)K_0\ra=16\pi\beta_b(\delta_{il}\delta_{kj}-\delta_{ij}\delta_{kl})\partial_k\partial_l[\delta(x_3-y_3)\delta(x_4-y_4)]\delta(x_1)\delta(x_2)\delta(y_1)\delta(y_2)\, .
\lb{2.12}
\ee
By means of relation (\ref{2}) (and using (\ref{1.5}) and (\ref{2.9})) this coincides precisely with (\ref{2.7}). The two methods thus give the same result for the 2-point function of energy momentum tensor
on a conical defect.

Let us again comment on a possible contribution of the anomaly of type $B$. In 3-point function  (\ref{2.8})  this contribution is given by function ${\cal A}^F_{\sigma\rho,\alpha\beta}(x-y,x-z)$   which was analyzed in  \cite{Osborn:1993cr}. We do not need its exact form however.
An important property of this term is that ${\cal A}^F_{\sigma\sigma,\alpha\beta}(x-y,x-z)=0$. Therefore, this term does not make any contribution to a 3-point function
where at least two traces of energy momentum tensor are present. Respectively, it does not make any contribution to 2-point function   $\la T(x)T(y)\ra_{{\cal C}_\alpha}$.
This is in agreement with our earlier discussion in section 4.1.

\section{ Dimension $d=2n$}
\setcounter{equation}0
\subsection{General formula}
The above consideration can be generalized to arbitrary  even dimension $d=2n$.  
In this section we shall present a derivation  based on a generalization of variation formula (\ref{2.2}),
\be
&&\la T_{\mu_1\nu_1}(x_1)\, ..\, T_{\mu_{n-1}\nu_{n-1}}(x_{n-1})T(x)\ra=\nonumber \\
&&\frac{2}{\sqrt{g(x_1)}} \frac{\delta}{\delta g^{\mu_1\nu_1}(x_1)}\, ..\, \frac{2}{\sqrt{g(x_{n-1})}} \frac{\delta}{\delta g^{\mu_{n-1}\nu_{n-1}}(x_{n-1})}\, \la T(x)\ra\, .
\lb{3.1}
\ee
The trace anomaly in dimension $d=2n$ 
\be
\la T(x)\ra=(-1)^{n+1}\,\frac{a_{2n}}{2^{2n}\, n!}\, E_{2n}(x)+..\, ,
\lb{3.2}
\ee
where we keep only the anomaly of type $A$ and neglect any other contribution. $E_{2n}$ is the Euler density
\be
E_{2n}(x)= \epsilon_{\mu_1\mu_2..\mu_{2n-1}\mu_{2n}}\epsilon^{\nu_1\nu_2..\nu_{2n-1}\nu_{2n}}R^{\mu_1\mu_2}_{\ \ \nu_1\nu_2}..
R^{\mu_{2n-1}\mu_{2n}}_{\ \ \nu_{2n-1}\nu_{2n}}\, .
\lb{3.3}
\ee
As before, we choose directions $1$ and $2$  to be orthogonal to the surface $\Sigma$  and $3,..,d$ to be parallel to the surface.
Respectively, we shall use notations in which  $i_k$ and $j_k$ take values $3,..,d$. The Euler density considered on a conical 
space was evaluated in \cite{Fursaev:1995ef},
\be
E_{2n}(x)|_{{\cal C}_\alpha}=8\pi(1-\alpha)n\delta_\Sigma\, \epsilon_{i_1..i_{2n-2}}\epsilon^{j_1..j_{2n-2}}R^{i_1i_2}_{\ \ j_1j_2}..R^{i_{2n-3}i_{2n-2}}_{\ \ j_{2n-3}j_{2n-2}}\, ,
\lb{3.4}
\ee
where $\delta_\Sigma$ is delta-function of variable $x$ which has support on surface $\Sigma$.
We notice that in this expression all indexes take values $3,..,d$ and  the Riemann tensor is in fact 
the intrinsic curvature of the surface $\Sigma$. Here we are interested in small variations over the flat metric so that the Riemann tensor takes the form
(\ref{2.4}). The calculation of variations (\ref{3.1}) is now straightforward. After some algebra we find
\be
&&\la T_{\mu_1\nu_1}(x_1)..T_{\mu_{n-1}\nu_{n-1}}(x_{n-1})T(x)\ra_{{\cal C}_\alpha}=2\pi(1-\alpha)\, a_{2n}\, \delta_\Sigma\nonumber \\
&&(-1)^{n+1} \epsilon^{\mu_1i_1..\mu_{n-1}i_{n-1}}\,
\epsilon^{\nu_1j_1..\nu_{n-1}j_{n-1}}\partial_{i_1}\partial_{j_1}\delta(x_1-x)..\partial_{i_{n-1}}\partial_{j_{n-1}}\delta(x_{n-1}-x)+ \mu_k\leftrightarrow\nu_k\, .
\lb{3.5}
\ee
This correlation function is non-vanishing for indexes $\mu_k,\, \nu_k$ taking values $3,..,d$.

As an application of this general formula   we consider

\subsection{ Example: $d=6$} 
In this case general formula (\ref{3.5}) reduces to
\be
&&\la T_{\mu_1\nu_1}(x_1)T_{\mu_2\nu_2}(x_2)T(x)\ra_{{\cal C}_\alpha}=\nonumber \\
&&2\pi(1-\alpha)a_6\, \delta_\Sigma \, \epsilon_{\mu_1\mu_2}^{\ \ \  \ \ i_1i_2}\epsilon_{\nu_1\nu_2}^{\ \ \ \ \ j_1j_2}\,
\partial_{i_1}\partial_{j_1}\delta(x_1-x)\partial_{i_2}\partial_{j_2}\delta(x_2-x)+\mu_k\leftrightarrow\nu_k\, .
\lb{3.6}
\ee
In particular for the correlation function of traces we obtain
\be
&&\la T(x_1)T(x_2)T(x)\ra_{{\cal C}_\alpha}=\nonumber \\
&&16\pi(1-\alpha)a_6\, \delta_\Sigma\, \left(\partial^2\delta(x_1-x)\partial^2\delta(x_2-x)-\partial_i\partial_j\delta(x_1-x)\partial^i\partial^j\delta(x_2-x)\right)\, ,
\lb{3.7}
\ee
where $\partial^2=\partial^i\partial_i$ is the Laplace operator on surface $\Sigma$.

\subsection{No contribution from $B$-anomaly}
Let us discuss a possible contribution of the anomaly of type B generally present in (\ref{3.2}). This anomaly, let us denote it by  $I$, is constructed from the Weyl tensor and its derivatives. Being considered on spacetime with a conical singularity this  gives
\be
I|_{{\cal C}_\alpha}=2\pi(1-\alpha){\cal J}\delta_\Sigma\, ,
\lb{3.8}
\ee
where ${\cal J}$ is conformal invariant constructed from projections of Weyl tensor and its derivatives on the transverse subspace.
Consider now the $n$-point correlation function 
\noindent $\la T(x_1)...T(x_{n-1})T(x)\ra$. It is obtained by taking traces in variation formula (\ref{3.1}). 
The respective contribution due to anomaly of type B is obtained by varying $(n-1)$ times equation (\ref{3.8}). By construction, it will necessarily contain at least one
variation of the Weyl tensor. By conformal invariance this  variation vanishes. We conclude that there is no contribution from anomaly of type B  to the $n$-point correlation
function of traces of energy momentum tensor. On the other hand, the respective contribution due to anomaly of type A is always present. It is easily seen by taking traces in 
equation (\ref{3.5}).

\section{Conclusion}
It is generally believed that in flat 4-dimensional spacetime the a-charge in the trace anomaly appears in  correlation functions
of a CFT energy momentum tensor starting with 3-point function and higher.
In this note we show that in the presence of a co-dimension two defect the situation is different and the a-charge
shows up already in a 2-point function.  This fact can be used to detect the a-charge either in a cosmic string spacetime or
in entanglement entropy where a conical defect appears as an intermediate technical trick. 
We generalize this observation for any even dimension $d=2n$ and give a general formula for a $n$-point correlation function
of energy momentum tensor on  a conical defect.  

Our results in this note remove the obstacles for the use of 2-point functions on conical defects
in proving the a-theorem in four dimensions. An idea of such a proof was outlined in \cite{SS} for the dilatonic contribution to entanglement entropy.
This and other ideas, such as presented in \cite{Rosenhaus:2014nha}, may be helpful in simplifying the existing proof \cite{Komargodski:2011vj}
and in exploring the new interesting directions.

\section*{Acknowledgements} 
I would like to thank Misha Smolkin for many helpful discussions. The kind hospitality
of the Theory Division at CERN and the Yukawa Institute for Theoretical Physics (Kyoto)  during the project is greatly acknowledged.


\begin{thebibliography}{999}

{\frenchspacing \parskip=2mm


\bibitem{SS} 
  S.~N.~Solodukhin,
  ``The a-theorem and entanglement entropy,''
  arXiv:1304.4411 [hep-th].

 
 
  
  
\bibitem{Capper:1974ic} 
  D.~M.~Capper and M.~J.~Duff,
  ``Trace anomalies in dimensional regularization,''
  Nuovo Cim.\ A {\bf 23}, 173 (1974).\\
   S.~Deser, M.~J.~Duff and C.~J.~Isham,
  ``Nonlocal Conformal Anomalies,''
  Nucl.\ Phys.\ B {\bf 111}, 45 (1976).\\
  M.~J.~Duff,
  ``Twenty years of the Weyl anomaly,''
  Class.\ Quant.\ Grav.\  {\bf 11}, 1387 (1994)
  [hep-th/9308075].


\bibitem{Deser:1993yx} 
  S.~Deser and A.~Schwimmer,
  ``Geometric classification of conformal anomalies in arbitrary dimensions,''
  Phys.\ Lett.\ B {\bf 309}, 279 (1993)
  [hep-th/9302047].
 
\bibitem{Cardy:1988cwa} 
  J.~L.~Cardy,
  ``Is There a c Theorem in Four-Dimensions?,''
  Phys.\ Lett.\ B {\bf 215}, 749 (1988).
 
\bibitem{Komargodski:2011vj} 
  Z.~Komargodski and A.~Schwimmer,
  ``On Renormalization Group Flows in Four Dimensions,''
  JHEP {\bf 1112}, 099 (2011)
  [arXiv:1107.3987 [hep-th]].\\
   Z.~Komargodski,
  ``The Constraints of Conformal Symmetry on RG Flows,''
  JHEP {\bf 1207}, 069 (2012)
  [arXiv:1112.4538 [hep-th]].
  
 
\bibitem{Osborn:1993cr} 
  H.~Osborn and A.~C.~Petkou,
  ``Implications of conformal invariance in field theories for general dimensions,''
  Annals Phys.\  {\bf 231}, 311 (1994)
  [hep-th/9307010].


\bibitem{Solodukhin:2008dh} 
  S.~N.~Solodukhin,
  ``Entanglement entropy, conformal invariance and extrinsic geometry,''
  Phys.\ Lett.\ B {\bf 665}, 305 (2008)
  [arXiv:0802.3117 [hep-th]].\\
  H.~Casini, M.~Huerta and R.~C.~Myers,
  ``Towards a derivation of holographic entanglement entropy,''
  JHEP {\bf 1105} (2011) 036
  [arXiv:1102.0440 [hep-th]].


\bibitem{EC}
H.~Casini and M.~Huerta,
  Phys.\ Lett.\ B {\bf 600}, 142 (2004)
  [hep-th/0405111].\\
R.~C.~Myers and A.~Sinha,
  ``Seeing a c-theorem with holography,''
  Phys.\ Rev.\ D {\bf 82}, 046006 (2010)
  [arXiv:1006.1263 [hep-th]].\\
  R.~C.~Myers and A.~Sinha,
  ``Holographic c-theorems in arbitrary dimensions,''
  JHEP {\bf 1101}, 125 (2011)
  [arXiv:1011.5819 [hep-th]].


\bibitem{Callan:1994py} 
  C.~G.~Callan, Jr. and F.~Wilczek,
  ``On geometric entropy,''
  Phys.\ Lett.\ B {\bf 333}, 55 (1994)
  [hep-th/9401072].

\bibitem{EE}
S.~Ryu and T.~Takayanagi,
  ``Aspects of Holographic Entanglement Entropy,''
  JHEP {\bf 0608}, 045 (2006)
  [hep-th/0605073].\\
  S.~N.~Solodukhin,
  ``Entanglement entropy of black holes,''
  Living Rev.\ Rel.\  {\bf 14}, 8 (2011)
  [arXiv:1104.3712 [hep-th]].
 
 
\bibitem{Smolkin:2014hba} 
  M.~Smolkin and S.~N.~Solodukhin,
  ``Correlation functions on conical defects,''
  arXiv:1406.2512 [hep-th].
 

\bibitem{Fursaev:1995ef} 
  D.~V.~Fursaev and S.~N.~Solodukhin,
  ``On the description of the Riemannian geometry in the presence of conical defects,''
  Phys.\ Rev.\ D {\bf 52}, 2133 (1995)
  [hep-th/9501127].

\bibitem{Cappelli:1990yc} 
  A.~Cappelli, D.~Friedan and J.~I.~Latorre,
  ``C theorem and spectral representation,''
  Nucl.\ Phys.\ B {\bf 352}, 616 (1991).
  
\bibitem{Rosenhaus:2014nha} 
  V.~Rosenhaus and M.~Smolkin,
  ``Entanglement Entropy Flow and the Ward Identity,''
  arXiv:1406.2716 [hep-th].
 
}


\end{thebibliography}
\end{document}